# Energy and Isotope Dependence of Neutron Multiplicity Distributions

J. P. Lestone

Los Alamos National Laboratory, Applied Physics Division, Los Alamos, New Mexico 87545

## Abstract

Fission neutron multiplicity distributions are known to be well reproduced by simple Gaussian distributions. Many previous evaluations of multiplicity distributions have adjusted the widths of Gaussian distributions to best fit the measured multiplicity distributions $P_n$. However, many observables do not depend on the detailed shape of $P_n$, but depend on the first three factorial moments of the distributions. In the present evaluation, the widths of Gaussians are adjusted to fit the measured 2$^{nd}$ and 3$^{rd}$ factorial moments. The relationships between the first three factorial moments are estimated assuming that the widths of the multiplicity distributions are independent of the initial excitation energy of the fissioning system. These simple calculations are in good agreement with experimental neutron induced fission data up to an incoming neutron energy of 10 MeV.

## I. INTRODUCTION

The measurement of the mean number of neutrons from a given type of fission event is relatively straightforward and has been performed for a wide range of fission reactions including neutron induced fission with thermal and fast neutrons [1]. For a given type of fission event, there is a probability $P_n$ that $n$ neutrons are emitted. This distribution is generally called the neutron multiplicity distribution. The measurement of neutron multiplicity distributions requires a large neutron detection apparatus with very high fission neutron detection efficiency. Spontaneous fission and thermal neutron induced fission neutron multiplicity distributions have been measured for a large number of heavy nuclei [2-5]. The high neutron detector efficiency needed to make neutron multiplicity distribution measurements, causes significant background problems in the case of non-thermal neutron induced reactions. This is because the incident neutrons can scatter from fission chambers and shielding materials into the neutron detector. The measurement of neutron multiplicity distributions in non-thermal neutron induced fission reactions is thus very difficult. We are aware of only a single set of such measurements on $^{235}$U, $^{238}$U, and $^{239}$Pu in the incident neutron energy range from 1 to 15 MeV [6].

Detailed knowledge of the neutron multiplicity distributions is required by non-destructive assay (NDA) techniques that infer the spontaneous fission rate, alpha-particle induced neutron emission, and induced fission rate from measured neutron multiplicity distributions from unknown samples [7]. These techniques are generally very accurate, in part, because the multiplication is usually small and often only needed to make corrections to extract the spontaneous fission rate which is usually the observable of most interest. However, for samples containing significant amounts of fissile material it becomes increasingly important to have a detailed knowledge of the fast neutron induced neutron multiplicity distributions.

## II. MEASURED $P_n$ FOR FAST NEUTRON INDUCED REACTIONS

Neutron multiplicities for fast neutron induced fission of $^{235}$U, $^{238}$U, and $^{239}$Pu are tabulated in Tables I, II, and III. These data were manually read from figures 10-15 in ref [6]. The $P_7$ values were estimated assuming the $P_n$ sum to 1 and the $P_n$ for all $n > 7$ are negligible. The mean ($v_1$) of the neutron multiplicity distributions given in Tables I-III are typically within ~1% of the measured mean neutron multiplicities. Despite the large uncertainty in the individual $P_n$ values, the 2$^{nd}$ ($v_2$) and 3$^{rd}$ ($v_3$) factorial moments ($\bar{v}_i = \sum P_n n!/(n-i)!$) of the neutron multiplicity distributions vary reasonably smoothly with the first moment ($v_1$) (see figures 1-3). This suggests that the uncertainty in the factorial moments is significantly less than the uncertainty of the individual $P_n$ values and that these measurements have accurately determined the relationships between the first three factorial moments of the multiplicity distributions.

Table I. Neutron multiplicity distributions $P_n$ for fast neutron induced fission of $^{235}$U [6]. Also given are the measured mean multiplicities, $\bar{\nu}$ [6], and the first three factorial moments of the given $P_n$.

| $\bar{\nu}$ | $P_0$ | $P_1$ | $P_2$ | $P_3$ | $P_4$ | $P_5$ | $P_6$ | $P_7$ | $\nu_1$ | $\nu_2$ | $\nu_3$ |
|---|---|---|---|---|---|---|---|---|---|---|---|
| 2.55 | 0.0216 | 0.163 | 0.304 | 0.326 | 0.153 | 0.039 | 0.001 | 0.00 | 2.56 | 5.21 | 8.1 |
| 2.63 | 0.0190 | 0.140 | 0.307 | 0.316 | 0.164 | 0.043 | 0.002 | 0.00 | 2.59 | 5.40 | 8.7 |
| 2.68 | 0.0155 | 0.133 | 0.301 | 0.327 | 0.181 | 0.046 | 0.006 | 0.00 | 2.71 | 5.84 | 9.8 |
| 2.75 | 0.0136 | 0.118 | 0.299 | 0.326 | 0.194 | 0.053 | 0.007 | 0.00 | 2.78 | 6.15 | 10.6 |
| 2.81 | 0.0150 | 0.101 | 0.300 | 0.326 | 0.197 | 0.067 | 0.002 | 0.00 | 2.81 | 6.32 | 10.9 |
| 2.89 | 0.0100 | 0.091 | 0.290 | 0.312 | 0.233 | 0.053 | 0.017 | 0.00 | 2.91 | 6.82 | 12.7 |
| 3.04 | 0.0070 | 0.077 | 0.256 | 0.333 | 0.229 | 0.096 | 0.007 | 0.00 | 3.03 | 7.39 | 14.1 |
| 3.25 | 0.0050 | 0.060 | 0.202 | 0.335 | 0.264 | 0.111 | 0.028 | 0.00 | 3.25 | 8.64 | 18.4 |
| 3.42 | 0.0020 | 0.032 | 0.178 | 0.332 | 0.298 | 0.129 | 0.034 | 0.00 | 3.43 | 9.52 | 21.0 |
| 3.43 | 0.0024 | 0.041 | 0.174 | 0.329 | 0.278 | 0.140 | 0.037 | 0.00 | 3.41 | 9.57 | 21.5 |
| 3.54 | 0.0017 | 0.029 | 0.153 | 0.323 | 0.312 | 0.143 | 0.042 | 0.00 | 3.52 | 10.11 | 23.0 |
| 3.59 | 0.0014 | 0.018 | 0.164 | 0.292 | 0.333 | 0.154 | 0.040 | 0.00 | 3.56 | 10.36 | 23.8 |
| 3.67 | -0.0004 | 0.024 | 0.133 | 0.307 | 0.323 | 0.159 | 0.054 | 0.00 | 3.62 | 10.78 | 25.6 |
| 3.74 | 0.0016 | 0.016 | 0.128 | 0.299 | 0.319 | 0.176 | 0.061 | 0.00 | 3.69 | 11.23 | 27.3 |
| 3.81 | -0.0004 | 0.012 | 0.118 | 0.284 | 0.326 | 0.199 | 0.050 | 0.01 | 3.77 | 11.75 | 29.6 |
| 3.87 | 0.0001 | 0.007 | 0.113 | 0.278 | 0.332 | 0.197 | 0.073 | 0.00 | 3.82 | 12.01 | 30.2 |
| 3.88 | -0.0002 | 0.012 | 0.104 | 0.270 | 0.324 | 0.218 | 0.054 | 0.02 | 3.88 | 12.54 | 33.2 |
| 3.97 | 0.0011 | 0.014 | 0.093 | 0.270 | 0.328 | 0.222 | 0.078 | 0.00 | 3.90 | 12.52 | 32.2 |
| 3.98 | 0.0009 | 0.007 | 0.086 | 0.267 | 0.332 | 0.184 | 0.112 | 0.01 | 3.97 | 13.22 | 36.2 |
| 4.09 | 0.0007 | 0.007 | 0.091 | 0.227 | 0.337 | 0.234 | 0.062 | 0.04 | 4.04 | 13.81 | 39.3 |
| 4.13 | 0.0001 | 0.008 | 0.070 | 0.243 | 0.312 | 0.260 | 0.071 | 0.04 | 4.13 | 14.35 | 41.5 |
| 4.21 | 0.0005 | 0.004 | 0.071 | 0.207 | 0.342 | 0.240 | 0.116 | 0.02 | 4.17 | 14.61 | 42.0 |
| 4.28 | 0.0001 | 0.002 | 0.066 | 0.206 | 0.343 | 0.250 | 0.104 | 0.03 | 4.21 | 14.86 | 43.2 |
| 4.35 | 0.0012 | 0.000 | 0.048 | 0.221 | 0.289 | 0.280 | 0.097 | 0.06 | 4.32 | 15.92 | 49.3 |
| 4.41 | 0.0000 | 0.003 | 0.044 | 0.186 | 0.280 | 0.377 | 0.018 | 0.10 | 4.46 | 16.84 | 53.6 |
| 4.49 | 0.0000 | 0.009 | 0.022 | 0.179 | 0.310 | 0.310 | 0.112 | 0.06 | 4.47 | 16.92 | 53.2 |
| 4.53 | 0.0000 | 0.003 | 0.048 | 0.146 | 0.338 | 0.248 | 0.193 | 0.03 | 4.50 | 17.04 | 53.3 |

## III. MODELING NEUTRON MULTIPLICITY DISTRIBUTIONS

Measured neutron multiplicity distributions have been previously parameterized using Gaussian distributions [8], and truncated renormalized single and double Gaussians [9]. In the present paper the relationship between the first three factorial moments is estimated assuming neutron multiplicity distributions are Gaussian [8] and that the widths of these Gaussians are independent of incoming neutron energy. By comparing the calculated relationship between the factorial moments to the corresponding experimental results [6] the validity of a fixed width as a function of incoming neutron energy can be tested.

Fig. 4 shows measured neutron multiplicity distributions for thermal-neutron induced fission of $^{235}$U and $^{239}$Pu. Neutron multiplicity distributions can be reasonably well represented by [8]

$$P_0 = \frac{1}{\sqrt{2\pi\sigma^2}} \int_{-\infty}^{1/2} \exp(\frac{-(x-\bar{\nu}+b)^2}{2\sigma^2}) \, dx, \text{ and}$$

$$P_{n\neq 0} = \frac{1}{\sqrt{2\pi\sigma^2}} \int_{n-1/2}^{n+1/2} \exp(\frac{-(x-\bar{\nu}+b)^2}{2\sigma^2}) \, dx, (1)$$

where $\bar{\nu}$ is the mean multiplicity, $b$ is a small adjustment to make the mean equal to $\bar{\nu}$, and $\sigma$ is the root-mean-square width. To determined the value of $\sigma$ from experimental data many authors have minimized the chi-squared

$$\chi^2(\sigma) = \sum_n \left[\frac{P_n^{\exp} - P_n(\sigma)}{\Delta P_n^{\exp}}\right]^2, \quad (2)$$

Table II. As for Table I, but for $^{238}$U [6].

| $\bar{\nu}$ | $P_0$ | $P_1$ | $P_2$ | $P_3$ | $P_4$ | $P_5$ | $P_6$ | $P_7$ | $\nu_1$ | $\nu_2$ | $\nu_3$ |
|---|---|---|---|---|---|---|---|---|---|---|---|
| 2.45 | 0.0222 | 0.200 | 0.306 | 0.307 | 0.136 | 0.035 | 0.004 | 0.00 | 2.48 | 4.91 | 7.7 |
| 2.60 | 0.0270 | 0.154 | 0.283 | 0.325 | 0.180 | 0.031 | 0.009 | 0.00 | 2.62 | 5.57 | 9.2 |
| 2.63 | 0.0184 | 0.144 | 0.302 | 0.332 | 0.158 | 0.043 | 0.013 | 0.00 | 2.67 | 5.74 | 9.9 |
| 2.68 | 0.0117 | 0.143 | 0.310 | 0.300 | 0.184 | 0.061 | 0.001 | 0.00 | 2.71 | 5.88 | 10.0 |
| 2.80 | 0.0176 | 0.114 | 0.283 | 0.315 | 0.213 | 0.056 | 0.006 | 0.00 | 2.79 | 6.31 | 11.1 |
| 2.88 | 0.0156 | 0.095 | 0.267 | 0.344 | 0.200 | 0.072 | 0.013 | 0.00 | 2.90 | 6.83 | 12.7 |
| 3.07 | 0.0890 | 0.069 | 0.230 | 0.359 | 0.235 | 0.086 | 0.017 | 0.00 | 3.08 | 7.66 | 15.0 |
| 3.23 | 0.0027 | 0.054 | 0.222 | 0.330 | 0.265 | 0.100 | 0.027 | 0.00 | 3.21 | 8.41 | 17.6 |
| 3.39 | -0.0280 | 0.041 | 0.207 | 0.278 | 0.334 | 0.109 | 0.035 | 0.00 | 3.38 | 9.32 | 20.4 |
| 3.40 | -0.0140 | 0.039 | 0.172 | 0.340 | 0.294 | 0.126 | 0.044 | 0.00 | 3.47 | 9.75 | 21.9 |
| 3.43 | 0.0140 | 0.035 | 0.170 | 0.337 | 0.291 | 0.143 | 0.018 | 0.00 | 3.37 | 9.25 | 19.7 |
| 3.55 | 0.0040 | 0.027 | 0.151 | 0.343 | 0.300 | 0.119 | 0.073 | 0.00 | 3.59 | 10.53 | 25.2 |
| 3.59 | 0.0040 | 0.017 | 0.150 | 0.315 | 0.315 | 0.147 | 0.054 | 0.00 | 3.58 | 10.53 | 24.8 |
| 3.69 | 0.0090 | 0.021 | 0.118 | 0.320 | 0.306 | 0.179 | 0.042 | 0.00 | 3.59 | 10.67 | 25.0 |
| 3.76 | 0.0070 | 0.012 | 0.133 | 0.276 | 0.331 | 0.186 | 0.054 | 0.00 | 3.68 | 11.23 | 27.2 |
| 3.80 | -0.0150 | 0.019 | 0.099 | 0.311 | 0.323 | 0.178 | 0.074 | 0.01 | 3.85 | 12.14 | 31.3 |
| 3.86 | -0.0190 | 0.010 | 0.110 | 0.274 | 0.344 | 0.204 | 0.052 | 0.03 | 3.94 | 12.68 | 33.6 |
| 3.91 | 0.0020 | 0.018 | 0.087 | 0.300 | 0.304 | 0.215 | 0.072 | 0.00 | 3.83 | 12.17 | 31.1 |
| 3.98 | 0.0120 | 0.015 | 0.083 | 0.264 | 0.325 | 0.198 | 0.111 | 0.00 | 3.93 | 12.94 | 34.6 |
| 4.07 | 0.0070 | 0.011 | 0.075 | 0.263 | 0.312 | 0.213 | 0.118 | 0.00 | 3.97 | 13.27 | 36.0 |
| 4.14 | 0.0140 | 0.009 | 0.066 | 0.250 | 0.298 | 0.236 | 0.146 | 0.00 | 4.14 | 14.31 | 40.3 |
| 4.20 | 0.0020 | 0.005 | 0.083 | 0.165 | 0.329 | 0.277 | 0.143 | 0.00 | 4.23 | 14.93 | 42.7 |
| 4.29 | 0.0170 | 0.010 | 0.043 | 0.257 | 0.270 | 0.292 | 0.081 | 0.03 | 4.10 | 14.40 | 41.6 |
| 4.34 | -0.0030 | 0.011 | 0.056 | 0.239 | 0.307 | 0.275 | 0.110 | 0.00 | 4.10 | 14.03 | 38.5 |
| 4.44 | 0.0000 | 0.005 | 0.063 | 0.148 | 0.257 | 0.410 | 0.060 | 0.06 | 4.43 | 16.62 | 51.5 |
| 4.49 | -0.0050 | 0.009 | 0.037 | 0.194 | 0.213 | 0.406 | 0.081 | 0.07 | 4.49 | 17.07 | 54.0 |
| 4.50 | 0.0130 | 0.009 | 0.029 | 0.189 | 0.289 | 0.282 | 0.184 | 0.00 | 4.30 | 15.82 | 47.1 |

where $\Delta P_n^{\text{exp}}$ is the uncertainty in the experimentally measured multiplicity distribution $P_n^{\text{exp}}$. The factorial moments of the neutron multiplicity distribution ($\bar{\nu}_i = \sum P_n n!/(n-i)!$) emitted by a multiplying sample can be expressed as a function of the factorial moments for spontaneous and induced fission [10]. Therefore, for many applications it is not necessary to know the details of the neutron multiplicity distribution but more important to know the corresponding first three factorial moments. In the present paper we fit the measured factorial moments instead of the details of the shape of the multiplicity distribution by minimizing the chi-squared

$$\chi^2(\sigma) = \sum_{i=2}^{3}\left[\frac{\nu_i(P_n^{\text{exp}}) - \nu_i(P_n(\sigma))}{\Delta \nu_i^{\text{exp}}}\right]^2 . \quad (3)$$

Determining the uncertainties in the experimental 2$^{\text{nd}}$ and 3$^{\text{rd}}$ moments is not a straightforward task and would require a detailed knowledge of the correlations between the measured $P_n^{\text{exp}}$ at different $n$. For simplicity, the relative uncertainty of the 3$^{\text{rd}}$ moments is assumed to be twice the relative uncertainty of the 2$^{\text{nd}}$ moment. Despite the change in emphasis from the detailed shape to the moments of the distributions, the inferred widths are little changed. However, by minimizing the chi-squared in Eq. (3) the inferred widths are guaranteed to be in reasonable agreement with the measured 2$^{\text{nd}}$ and 3$^{\text{rd}}$ factorial moments. The open histograms in Fig. 4 are the Gaussian fits to the corresponding experimental data, obtained by minimizing the chi-squared in Eq. (3). Experimentally measured $P_n^{\text{exp}}$ and the corresponding Gaussian fits $P_n(\sigma)$ for thermal neutron induced fission of $^{235}$U and $^{239}$Pu are given in table IV along with the corresponding first three moments. These Gaussian distributions reproduce the experimental first three factorial moments to better than 0.6%. If instead, the chi-squared in Eq. (2) was used then the corresponding factorial moments can differ from the experimental values by as much as 10%.

If the root-mean-squared width of $P_n(\sigma)$ is assumed to be independent of incoming neutron kinetic energy then the relationship between different factorial moments is easily calculated as a function of $\bar{\nu}$. The solid curves in Figs. 1-3 show such calculations with $\sigma$ =1.088, 1.116, and 1.140. The value of $\sigma$ =1.116 for $^{238}$U(n,f) was obtained by adjusting $\sigma$ to fit the $\bar{\nu}$ <4.0 data shown in Fig. 2. Below $\bar{\nu}$ ~4.0 the solid curves in Figs 1-3 are in agreement with the data, and thus we can conclude that the fixed width assumption is reasonable up to an incoming neutron energy of $E_n$~10 MeV. Above $\bar{\nu}$ ~4.0 the solid curves in Figs 1-3 are systematically low, suggesting that the neutron multiplicity distributions are noticeably wider for $E_n$>10 MeV.

## IV. RECOMMENDATIONS FOR MONTE-CARLO NEUTRON-TRANSPORT CODES

Recommended values for the root-mean-squared width of multiplicity distributions for various fission reactions are given in Table V. These widths were determined by minimizing the chi-squared defined in Eq. (3). These root-mean-squared widths are summarized in Fig. 5.

For reactions not given in Table V, widths can be estimated using the line shown in Fig. 5. Notice that the horizontal axis in Fig. 5 is the mass of the initial compound system, i.e. $^{A-1}Z$(n,f) results are plotted at mass number $A$. For Fm isotopes heavier than $A$=254 [5] the widths are much larger and do not follow the trend shown in Fig. 5. This is because, for $A$>254, the mass distributions become increasingly symmetric because of the $N$=82 magic number. Based on the good agreement between the experimental data and the model calculations shown in Figs 1-3, the width, $\sigma$, can be assumed to be independent of incoming neutron energy for $E_n$<10 MeV. The mean neutron multiplicity, $\bar{\nu}$, and the shift parameter, $b$, must of course, be varied as a function of $E_n$. The mean neutron multiplicities as a function of $E_n$ have been well defined by experiment [1], and are already accurately placed into most neutron transport Monte-Carlo codes. The shift parameter, $b$, is a function of $\sigma$, and $\bar{\nu}$. For the range of widths appropriate to real fission reactions, $b$ is a simple function of $(\bar{\nu} + 0.5)/\sigma$ as given in Table VI.

Table III. As for Table I, but for $^{239}$Pu [6].

| $\bar{\nu}$ | $P_0$ | $P_1$ | $P_2$ | $P_3$ | $P_4$ | $P_5$ | $P_6$ | $P_7$ | $v_1$ | $v_2$ | $v_3$ |
|---|---|---|---|---|---|---|---|---|---|---|---|
| 3.07 | 0.0061 | 0.077 | 0.251 | 0.326 | 0.245 | 0.089 | 0.015 | 0.00 | 3.07 | 7.63 | 15.0 |
| 3.16 | 0.0054 | 0.065 | 0.241 | 0.314 | 0.268 | 0.095 | 0.020 | 0.00 | 3.16 | 8.08 | 16.4 |
| 3.23 | 0.0056 | 0.057 | 0.214 | 0.329 | 0.273 | 0.101 | 0.025 | 0.00 | 3.22 | 8.45 | 17.6 |
| 3.31 | 0.0051 | 0.051 | 0.203 | 0.320 | 0.272 | 0.122 | 0.034 | 0.00 | 3.32 | 9.05 | 19.8 |
| 3.38 | 0.0059 | 0.042 | 0.196 | 0.307 | 0.293 | 0.125 | 0.032 | 0.00 | 3.34 | 9.21 | 20.2 |
| 3.47 | 0.0030 | 0.043 | 0.157 | 0.339 | 0.285 | 0.149 | 0.033 | 0.00 | 3.46 | 9.74 | 21.8 |
| 3.65 | 0.0011 | 0.027 | 0.147 | 0.308 | 0.292 | 0.170 | 0.052 | 0.00 | 3.58 | 10.61 | 25.3 |
| 3.81 | 0.0011 | 0.030 | 0.109 | 0.275 | 0.340 | 0.155 | 0.080 | 0.01 | 3.76 | 11.87 | 30.8 |
| 3.96 | -0.0015 | 0.014 | 0.104 | 0.237 | 0.354 | 0.211 | 0.064 | 0.02 | 3.91 | 12.77 | 34.0 |
| 4.00 | 0.0015 | 0.018 | 0.092 | 0.234 | 0.343 | 0.224 | 0.062 | 0.03 | 3.94 | 13.09 | 35.8 |
| 4.03 | 0.0010 | 0.012 | 0.089 | 0.250 | 0.328 | 0.224 | 0.080 | 0.02 | 3.96 | 13.17 | 35.8 |
| 4.13 | -0.0004 | 0.012 | 0.075 | 0.237 | 0.341 | 0.219 | 0.098 | 0.02 | 4.05 | 13.74 | 38.3 |
| 4.19 | 0.0000 | 0.007 | 0.075 | 0.208 | 0.351 | 0.219 | 0.114 | 0.03 | 4.15 | 14.50 | 42.0 |
| 4.26 | 0.0000 | 0.006 | 0.067 | 0.209 | 0.313 | 0.267 | 0.108 | 0.03 | 4.21 | 14.98 | 44.0 |
| 4.34 | 0.0000 | 0.008 | 0.048 | 0.210 | 0.313 | 0.265 | 0.122 | 0.03 | 4.28 | 15.50 | 46.5 |
| 4.35 | -0.0006 | 0.004 | 0.052 | 0.204 | 0.310 | 0.272 | 0.123 | 0.04 | 4.30 | 15.65 | 47.1 |
| 4.41 | 0.0002 | 0.004 | 0.051 | 0.183 | 0.309 | 0.300 | 0.108 | 0.05 | 4.35 | 16.04 | 48.9 |
| 4.48 | -0.0004 | -0.001 | 0.056 | 0.161 | 0.321 | 0.281 | 0.130 | 0.05 | 4.43 | 16.63 | 52.1 |
| 4.56 | 0.0001 | 0.006 | 0.041 | 0.167 | 0.282 | 0.297 | 0.144 | 0.06 | 4.51 | 17.37 | 56.1 |
| 4.62 | -0.0005 | 0.004 | 0.036 | 0.154 | 0.296 | 0.263 | 0.201 | 0.05 | 4.57 | 17.81 | 57.8 |
| 4.70 | -0.0003 | -0.003 | 0.053 | 0.114 | 0.311 | 0.275 | 0.169 | 0.08 | 4.65 | 18.49 | 61.9 |
| 4.71 | 0.0006 | 0.004 | 0.035 | 0.100 | 0.373 | 0.241 | 0.294 | 0.00 | 4.84 | 18.79 | 59.3 |
| 4.83 | -0.0002 | 0.003 | 0.037 | 0.109 | 0.292 | 0.272 | 0.205 | 0.08 | 4.74 | 19.27 | 65.8 |
| 4.87 | 0.0009 | 0.004 | 0.080 | 0.136 | 0.279 | 0.229 | 0.300 | 0.00 | 4.63 | 17.90 | 57.3 |
| 4.96 | 0.0009 | 0.006 | -0.030 | 0.147 | 0.204 | 0.355 | 0.186 | 0.13 | 5.01 | 21.45 | 76.9 |
| 5.01 | -0.0005 | 0.001 | 0.027 | 0.109 | 0.181 | 0.359 | 0.166 | 0.16 | 4.99 | 21.59 | 79.2 |
| 5.05 | -0.0002 | 0.006 | 0.060 | 0.093 | 0.272 | 0.279 | 0.195 | 0.11 | 4.79 | 19.78 | 69.3 |

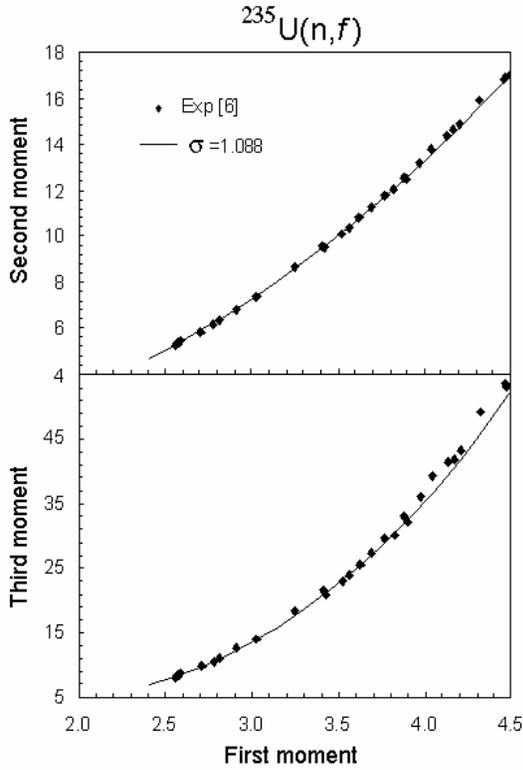

Fig. 1. The second and third factorial moments versus the first factorial moment for neutron induced fission of $^{235}$U. The small symbols show the experimental results of ref [6]. The solid lines are a model calculations assuming that the neutron multiplicity distributions are Gaussian [8] with a fixed width, $\sigma$ =1.088, independent of incoming neutron energy.

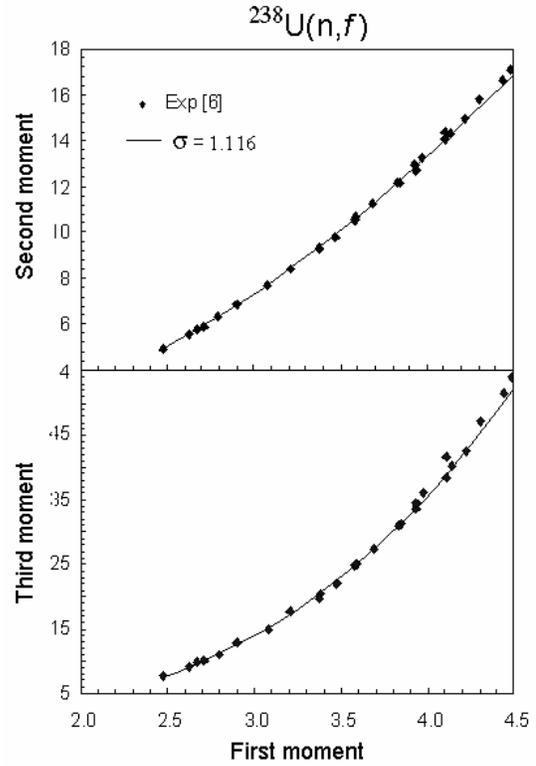

Fig. 2. As for Fig. 1 but for neutron induced fission of $^{238}$U and $\sigma$=1.116.

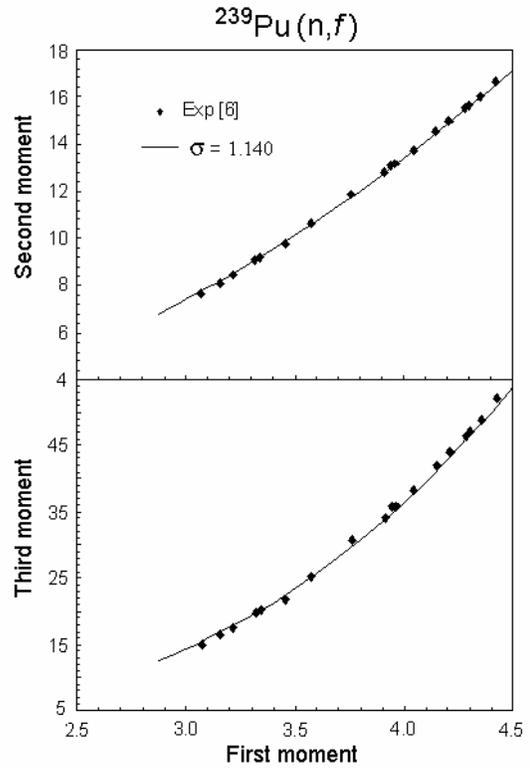

Fig. 3. As for Fig. 1 but for neutron induced fission of $^{239}$Pu and $\sigma$=1.140.

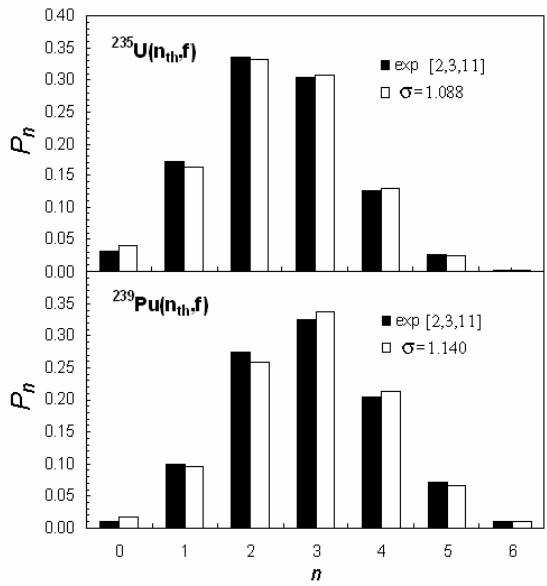

Fig. 4. Experimentally measured and Gaussian multiplicity distributions for thermal neutron induced fission of $^{235}$U and $^{239}$Pu.

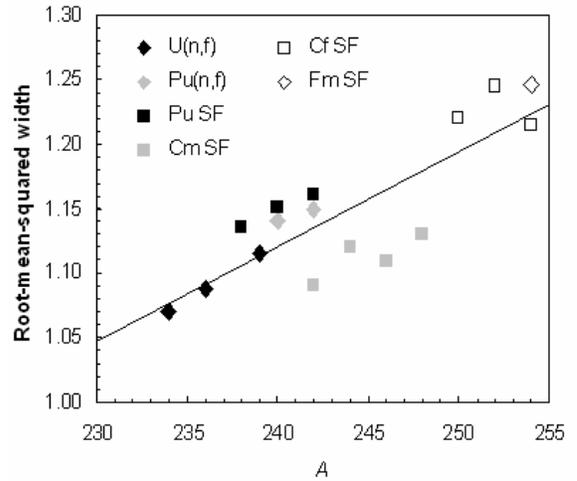

Fig. 5. Root-mean-squared width of neutron multiplicity distributions versus mass number of the fissioning system.

Table IV. Experimentally measured neutron multiplicity distributions, $P_n^{\exp}$ [2,3,11], the corresponding Gaussian fits, $P_n(\sigma)$, and the first three moments ($\nu_1$, $\nu_2$, and $\nu_3$) for thermal neutron induced fission of $^{235}$U and $^{239}$Pu.

| | $^{235}$U(n$_{th}$,f) | | $^{239}$Pu(n$_{th}$,f) | |
|---|---|---|---|---|
| $n$ | $P_n^{\exp}$ | $P_n(\sigma=1.088)$ | $P_n^{\exp}$ | $P_n(\sigma=1.140)$ |
| 0 | 0.0317±0.0015 | 0.0398 | 0.0109±0.0001 | 0.0187 |
| 1 | 0.1720±0.0014 | 0.1621 | 0.0995±0.0028 | 0.0955 |
| 2 | 0.3363±0.0031 | 0.3314 | 0.2750±0.0003 | 0.2576 |
| 3 | 0.3038±0.0004 | 0.3085 | 0.3270±0.0041 | 0.3372 |
| 4 | 0.1268±0.0036 | 0.1308 | 0.2045±0.0087 | 0.2144 |
| 5 | 0.0266±0.0026 | 0.0251 | 0.0728±0.0133 | 0.0661 |
| 6 | 0.0026±0.0009 | 0.0022 | 0.0097±0.0027 | 0.0099 |
| 7 | 0.0002±0.0001 | 0.0000 | 0.0006±0.0009 | 0.0007 |
| $\nu_1$ | 2.413 | 2.413 | 2.875 | 2.875 |
| $\nu_2$ | 4.635 | 4.655 | 6.738 | 6.761 |
| $\nu_3$ | 6.816 | 6.778 | 12.528 | 12.475 |

Table. V. Recommended Gaussian parameters for various neutron multiplicity distributions.

| Reaction | Data ref | $\sigma$ | $\bar{\nu}$ | $b$ |
|---|---|---|---|---|
| $^{233}$U(n$_{th}$,f) | 2,3,11 | 1.070 | 2.477 | 0.003 |
| $^{235}$U(n$_{th}$,f) | 2,3,11 | 1.088 | 2.413 | 0.004 |
| $^{238}$U(n$_{fast}$,f) | 6 | 1.116 | varying | varying |
| $^{239}$Pu(n$_{th}$,f) | 2,3,11 | 1.140 | 2.875 | 0.002 |
| $^{241}$Pu(n$_{th}$,f) | 2,3,11 | 1.150 | 2.929 | 0.002 |
| $^{238}$Pu SF[1] | 4,12 | 1.135 | 2.212 | 0.009 |
| $^{240}$Pu SF | 2 | 1.151 | 2.153 | 0.012 |
| $^{242}$Pu SF | 2 | 1.161 | 2.143 | 0.013 |
| $^{242}$Cm SF | 13 | 1.091 | 2.54 | 0.003 |
| $^{244}$Cm SF | 13 | 1.103 | 2.72 | 0.002 |
| $^{246}$Cm SF | 13 | 1.098 | 2.93 | 0.001 |
| $^{248}$Cm SF | 13 | 1.108 | 3.13 | 0.001 |
| $^{250}$Cf SF | 13 | 1.220 | 3.51 | 0.001 |
| $^{252}$Cf SF | 2 | 1.245 | 3.76 | 0.000 |
| $^{254}$Cf SF | 13 | 1.215 | 3.85 | 0.000 |
| $^{254}$Fm SF | 5 | 1.246 | 3.94 | 0.000 |

[1] SF : Spontaneous fission.

Table VI. The relationship between $\sigma$, $\bar{\nu}$, and $b$.

| $\dfrac{\bar{\nu}+0.5}{\sigma}$ | $b$ | $\dfrac{\bar{\nu}+0.5}{\sigma}$ | $b$ | $\dfrac{\bar{\nu}+0.5}{\sigma}$ | $b$ | $\dfrac{\bar{\nu}+0.5}{\sigma}$ | $b$ | $\dfrac{\bar{\nu}+0.5}{\sigma}$ | $b$ |
|---|---|---|---|---|---|---|---|---|---|
| 3.7227 | 0.0001 | 2.8199 | 0.0025 | 2.4075 | 0.0085 | 2.2097 | 0.0145 | 2.0743 | 0.0205 |
| 3.5455 | 0.0002 | 2.7114 | 0.0035 | 2.3673 | 0.0095 | 2.1841 | 0.0155 | 2.0552 | 0.0215 |
| 3.4379 | 0.0003 | 2.6281 | 0.0045 | 2.3307 | 0.0105 | 2.1599 | 0.0165 | 2.0369 | 0.0225 |
| 3.3597 | 0.0004 | 2.5600 | 0.0055 | 2.2971 | 0.0115 | 2.1369 | 0.0175 | 2.0192 | 0.0235 |
| 3.2980 | 0.0005 | 2.5023 | 0.0065 | 2.2659 | 0.0125 | 2.1151 | 0.0185 | 2.0022 | 0.0245 |
| 2.9785 | 0.0015 | 2.4521 | 0.0075 | 2.2369 | 0.0135 | 2.0943 | 0.0195 | 1.9857 | 0.0255 |